# Multi-TGDR: a regularization method for multi-class classification in microarray experiments

**Suyan Tian**[1,2§], **Mayte Suárez-Fariñas**[2,3§]

[1]Division of Clinical Epidemiology, First Hospital of the Jilin University, 71Xinmin Street, Changchun, Jilin, China. 130021
[2]Center for Clinical and Translational Science, [3]Laboratory for Investigative Dermatology, The Rockefeller University, 1230 York Avenue, New York, NY, USA 10065.

[§]Corresponding author

## Abstract

**Background**

With microarray technology becoming mature and popular, the selection and use of a small number of relevant genes for accurate classification of samples is a hot topic in the circles of biostatistics and bioinformatics. However, most of the developed algorithms lack the ability to handle multiple classes, which arguably a common application. Here, we propose an extension to an existing regularization algorithm called Threshold Gradient Descent Regularization (TGDR) to specifically tackle multi-class classification of microarray data. When there are several microarray experiments addressing the same/similar objectives, one option is to use meta-analysis version of TGDR (Meta-TGDR), which considers the classification task as combination of classifiers with the same structure/model while allowing the parameters to vary across studies. However, the original Meta-TGDR extension did not offer a solution to the prediction on independent samples. Here, we propose an explicit method to estimate the overall coefficients of the biomarkers selected by Meta-TGDR. This extension permits broader applicability and allows a comparison between the predictive performance of Meta-TGDR and TGDR using an independent testing set.

**Results**

Using real-world applications, we demonstrated the proposed multi-TGDR framework works well and the number of selected genes is less than the sum of all individualized binary TGDRs. Additionally, Meta-TGDR and TGDR on the batch-effect adjusted pooled data approximately provided same results. By adding Bagging procedure in each application, the stability and good predictive performance are warranted.

**Conclusions**

Compared with Meta-TGDR, TGDR is less computing time intensive, and requires no samples of all classes in each study. On the adjusted data, it has approximate same predictive performance with Meta-TGDR. Thus, it is highly recommended.

# Background

Biomarker discovery from high-dimensional data is a crucial problem with enormous applications in areas of biomedical research and translational medicine. Selecting a small number of relevant features (e.g., genes in transcriptomics profiles, SNPs in GWAs studies, and metabolites in metabolomics) to build a predictive model that can accurately classify samples by their diagnosis (e.g., diseased or health, different stages of one specific cancer) and prognosis (e.g., potential response to a given treatment, 5-year survival with a certain treatment) is an essential step towards personalized medicine. In bioinformatics, such a task is accomplished by a feature selection algorithm, which besides reducing over-fitting and improving classification accuracy, leads to small molecular signatures with manageable experimental verification and the potential design of cheap dedicated diagnostic and prognostic tools.

Among dozens to hundreds of proposed feature selection algorithms [1-3], the Threshold Gradient Decent Regularization (TGDR) proposed by Friedman and Popescu [4] stands out because of the elegant theory beneath them; easy to moderate programming for a well-trained statistician; good performance and biologically meaningful results in real-world applications. Ma and Huang[5] elegantly extended the TGDR to the case where expression data from several studies are combined. The proposed algorithm, the Meta Threshold Gradient Descent Regularization (Meta-TGDR), assumes that the same set of genes is selected on all studies, while allowing the $\beta$ coefficients to vary across studies, in a meta-analysis fashion. In their paper, they demonstrated that a better classification performance was achieved by using Meta-TGDR rather than by using TGDR on the combined data set.

However, both originally proposed TGDR and Meta-TGDR frameworks do not give the explicit definition or/and format on the multi-class classification where an observation needs be categorized into more than two classes. Additionally, Meta-TGDR[5] does not offer an overall predictive rule on an independent data set (testing samples), from a study not used in classifier training/estimation. The absence of such rule prevented Meta-TGDR from the evaluation of its performance on independent testing sets, and the comparison with TGDR in terms of predictive performance. Furthermore, it limited the use of the Meta-TGDR to real, clinical practice application precluding its use in personalized medicine, where a classifier is built for use in an extended population under variable laboratory setting.

In this paper, we specifically addressed the first issue by proposing a new framework, referred as to multi-TGDR, and the second issue by proposing an equation. Lastly, the results from Meta-TGDR, and TGDR on the pooled data were compared in terms of their predictive performance.

# Methods and Material

## The proposed extensions to TGDR and meta-TGDR

In order to establish the nomenclature to be used through the paper and to help the reader experience we start with a brief description of the Meta-TGDR as below. The interested readers are referred to [5] for more details on both TGDR and Meta-TGDR frameworks.

Let $Y^m = (Y_1^m,...,Y_{n_m}^m)$ be the indicator function (i.e., 0 if in the reference class, 1 otherwise) for each study m=1,..,M with $n_m$ subjects and $X^m = (X_1^m,...,X_{n_m}^m)$ the vector of $n_m \times D$ matrices representing the gene expression for each subject over the same set of D genes. The likelihood function for study *m* can be written as:

$$R^m(\beta^m) = \sum_{j=1}^{n_m} \left( Y_j^m \times (\beta_0^m + \beta^m X_j) - \log(1 + \exp(\beta_0^m + \beta^m X_j)) \right) \quad (1)$$

where $\beta_0^m$ and $\beta^m = (\beta_1^m,...,\beta_D^m)$ are the unknown intercept and expression-coefficients for each study s. The overall likelihood function can be written as $R(\beta)=R^1(\beta^1)+...+R^M(\beta^M)$ with $\beta = (\beta^1,...,\beta^M)$. Only the $\beta_i$s are subject to regularization.

Denote $\Delta v$ as the small positive increment (e.g., 0.01) as in ordinary gradient descent searching and $v_k = k \times \Delta v$ as the index for the point along the parameter path after *k* steps. Let $\beta(v_k)$ denote the parameter estimate of β corresponding to $v_k$. For a fixed threshold $0 \leq \tau \leq 1$, the Meta-TGDR algorithm can iterate on the following steps:

1. Initialize β(0)=0 and $v_0$=0.
2. With current estimate β,
    i) Compute the negative gradient matrix $g_{j,m}(v) = -\partial R^m(\beta^m)/\partial \beta_j^m$
    ii) Define the meta-gradient G(v) as a D-dimensional vector whose $j^{th}$-element is the sum of the gradient for each study; i.e., $G_j(v) = \sum_{m=1}^{M} g_{j,m}(v)$.
    iii) if $\max_j\{|G_j(v)|\} = 0$, stop the iteration.
3. Compute the threshold vector f(v) of length D, where the $j^{th}$ component of f(v):
$$f_j(v) = I(|G_j(v)| \geq \tau \times \max_l(|G_l(v)|))$$
4. Update β(v+Δv)= β(v) - Δv×g(v)×f(v) and update v by v+Δv , where the product of *f* and *g* is component-wise.
5. Steps 2-4 are iterated *k* times. Both τ and *k* are tuning parameters and determined by cross validation.

## *Multi-TGDR: Extension to multi-class classification*

Multiple-class classification is commonly encountered in real world, however, many proposed feature selection algorithms lack the valuable capacity of dealing with multi-class classification. In the original TGDR and Meta-TGDR framework, multi-class classification had been left untouched even though all authors claimed that such extension is very natural. Here, we propose an extension of the TGDR framework to multi-class cases.

In the multi-class scenario, the response variable $Y_i$ -representing the class membership for subject *i* - may take values 1,…,K, where *K* is the number of classes (K≥3). Propositions to tackle this problem using existing binary classifiers divided into two major types: "one-versus-the rest" where K binary classifiers were trained to distinguish the samples in a single class from the samples in all remaining classes, and "one-versus-another" where K(K-1)/2 classifiers were trained to distinguish the samples in a class from the samples in one remaining class. Many researchers [7-9] had demonstrated that one-versus-another schema offered better performance than one-versus-the rest did. Therefore, we compared our proposed framework with one-versus-another schema only.

The central idea of our extension is to replace the single indicator variable $Y_i$ for each sample for a set of K-1 variables $Y_{ik}$. The threshold function is then defined as the maximum along the set of local threshold functions, defined on the subspaces defined by parameters associated to each class. To our knowledge, multi-class TGDR has not been addressed; probably because it is more computationally demanding than binary TGDR.

Let $Y_{k1},…,Y_{kn}$ be the vector of indicators for class $k$ across subjects; i.e., $Y_{kj}$ is equal to 1 if the $j^{th}$ subject belongs to class k and zero otherwise. This vector is defined for each class k (k=1,…,K-1), so the $K^{th}$-class serves as the reference class. As before, let $X_1,…,X_n$ represent the gene expression values. After a simple algebraic manipulation，the log-likelihood function can be written as:

$$R(\beta) = \sum_{j=1}^{n}\left(\sum_{k=1}^{K-1} Y_{kj} \times (\beta_{k0} + \beta_k X_j) - \log(1 + \sum_{k=1}^{K-1} \exp(\beta_{k0} + \beta_k X_j))\right) \quad (2)$$

$\beta_{k0}$'s are unknown intercepts which would be not subject to regularization. $\beta_k = (\beta_{k1},…, \beta_{kD})$ are the corresponding coefficients for expression values for the same set of D genes for the comparison between class k and reference class K. Note, the dimension of all $\beta_k$ is restricted to be the same, but their magnitudes differ. Hence the same number of non-zero genes is used on all classes but their estimated values are different.

Let $\beta = \{\beta_{k0}, \beta_k\}_{k=1}^{K-1}$ denote the set of all parameters to be estimated in model 3, one can follow the binary TGDR procedure as detailed in [5, 10, 11] but introducing the following modification in the calculation of the threshold vector f(v) in step 3:

Here, $f_{ki}(v)$ represents the threshold vector of size D for class k (k=1,..,K-1),

$$f_{ki}(v) = I(|g_{ki}(v)|) \geq \tau \times \max_{l \in \beta_k}(|g_{kl}(v)|) \quad for \quad i \in \beta_k$$

Then, the $i^{th}$-gene specific element of threshold function f(v) will be obtained as:

$$f_i(v) = \max_k (f_{ki})$$

Thus, when one gene was selected in one comparison, it would appear in the rest comparisons but it may not differ significantly from zeros in those comparisons.

Although here we establish a unique $\tau$ tuning parameter, this assumption may be loosed so that $\tau$ can have different values for each class, which will allow different degree of regularization for different comparisons. Here, the proposed framework is referred to as multi-TGDR.

*Multi-class Meta-TGDR*

The Multi-TDGR can naturally be extended to a situation with multiples studies and multiple classes. The step 3 in multi-class TGDR is combined with the concept of meta-gradient of Meta-TGDR. That means $f_{ki}$ is defined on the meta-gradient instead of the regular gradient, i.e.,

$f_{ki}(v) = I(|G_{ki}(v)|) \geq \tau \times \max_{l \in \beta_k}(|G_{kl}(v)|)$ $for$ $i \in \beta_k$ and $f_i(v) = \max_k(f_{ki})$. Obviously, it is more time-consuming compared to multi-class TGDR. In this paper, this framework is referred as to Meta-multi-TGDR.

### *Prediction of new samples using Meta-TGDR*

In the Meta-TGDR, the estimated coefficients β = (β$_1$,…,β$_D$), corresponding to expression values for D genes selected, are different in each study. This raises a question of how to use these study-specific coefficients to perform an overall prediction in a new independent sample ("testing sample"), not previously used in the training/estimation stage. However [5] did not offer a solution to this issue, which preclude the evaluation of the performance of the Meta-TGDR (and its comparisons with TGDR) on independent "testing" samples. Here we extended their work by conducting a meta-regression to synthesize the results from Meta-TGDR and extrapolate the membership prediction to a sample from a new study. Under the Meta-TGDR settings, let $Z_{ij}$ be the estimated log odds for the $j^{th}$ sample in $i^{th}$ study using the estimated coefficients $\hat{\beta}^i$ (of length D) for each study (i.e., $Z_{ij} = X_j^i \hat{\beta}^i$). $Z_{ij}$ can be modeled as:

$$Z_{ij} = \mu_0 + \mu_1 X_{1j} + ... \mu_D X_{Dj} + \varepsilon_{ij} \qquad i = 1,...,M \quad j = 1,...,n_i \quad (3)$$

where μ$_1$,…, μ$_D$ represents the overall coefficient associated with gene i, $\varepsilon_{ij} \sim N(0, \sigma_i^2)$ and $\sigma_i^2$ is the within-study variance for the study *i*. Specification of equation 3 can be achieved by following a 3-step procedure with the first two steps obtain estimates of $\sigma_i^2$ using delta method[6] and step 3 calculates the posterior probabilities using the estimated coefficients obtained in step 2.

1. Consider $E(Y_{ij} | X_{ij}) = \exp(Z_{ij})/(1 + \exp(Z_{ij}))$, and $E(Y_i) = E(E(Y_{ij} | X_{ij}))$. For each study estimate the variance of Y in natural scale by:
$$S_i = \sum_{j=1}^{n_j} (Y_{ij} - E(Y_{ij} | X_{ij}))^2 / n_i$$
2. Estimate $\sigma_i^2$ using delta-method i.e.,
$$\hat{\sigma}_i^2 = S_i / (E(Y_i) \times (1 - E(Y_i)))^2$$
Once we have $\hat{\sigma}_i^2$, the overall estimated coefficients μs can be easily obtained by a weighted least square equation (similar to the one used in fixed-effect meta-analysis).
3. Finally, the overall μ$_i$'s estimated in step 2 is used to calculate the odd-ratio and the posterior class-membership probability for a new sample.

## Miscellaneous

### *Stabilization of the selected genes using Bootstrap aggregating*

In order to improve the stability and classification accuracy of multi-class TGDR, we applied Bootstrap aggregating (Bagging) to our classifier [12]. Given a training set of size n, bagging generates *m* new training sets, each of size n, by sampling subjects from the original training set with replacement. Then *m* multi-class TDGRs are conducted using the above *m* bootstrap samples and combined by voting. Bagging helps to protect over-fitting which usually exists in the classification setting.

*Evaluation of predictive performance*

The performance of a classifier is measured using traditional performance metrics over the training samples (training and cross-validation errors) and over the test samples (the predictive error). Since the membership probabilities for each sample can easily being obtained from TGDR algorithm we also used the generalized Brier score (GBS) proposed by Yeung et al [13] as a generalization of the Brier Score to a multi-class classification problem. Under the K class setting, where $Y_{ik}$ are indicator functions of class $k$ ($k$=0,1,…,$K$−1), let $\hat{p}_{ik}$ denote the predicted probability such that $Y_{ik}$=1. For easier interpretation and comparison of GBS score across different classification settings, we normalized the GBS by the sample size n as in [14] i.e., $\sum_{k=0}^{K-1}(Y_{ik}-\hat{p}_{ik})^2/2n$. By taking into consideration the magnitudes of predicted probabilities, GBS, can establish difference in performance of classifiers with same overall predictive error. The smaller the GBS value, which after normalization takes values in [0,1], the better a classifier performs.

*The Experimental Data and preprocessing procedures*

**Psoriasis:** Open-access data from 3 published studies [15-17] available under GEO accession numbers GSE14905, GSE13355, and GSE30999, respectively were used, including samples from Lesions (LS) and adjacent Non-Lesional (NL) skin form psoriasis patients and Normal skin from healthy patients. Details of these studies - using Hgu133plus2 Affymetrix chips- are given in [18].

**Lung cancer:** The lung cancer data sets included GSE10245, GSE18842, and GSE2109, all studies were performed on Affy HGU 133plus2 chips and publicly available on the GEO repository.

**Pre-processing procedures:** The raw Affymetrix data (CEL files) of both lung cancer and psoriasis data sets were downloaded from GEO repository and expression values were obtained using **FRMA** algorithm [19]. To pool data from different studies together and to address the batch effects from different experiments, COMBAT algorithm [20] was used to adjust on the combined expression values for these two combined data sets. For the lung cancer data, moderated F/t-tests (limma package) were conducted to identify differentially expressed genes (DEGs) with cutoffs for FDR and fold change as 0.05 and 2, respectively. When there are multiple probesets representing the same gene, the one with the largest F-value was chosen. The resulting 949 unique genes were fed into the downstream analysis. Note, for the TGDR algorithm there is no limit on the number of genes fed into the algorithm. However, we usually used a filtering procedure to rule out the non-informative genes before the classification. By doing so, a large amount of computing time can be saved with only partial set of genes put in classifiers; but with no or least loss on the potential biomarkers since almost all genes which have high probability to be biomarkers pass the filtering. For psoriasis data, the filtering steps taken (including SD, ICC and DEGs using meta-analysis method) were used by us previously and described in details there [18]. Similar to the lung cancer data, conducting those filtering steps is mainly for the purpose of saving the computing time. 2301 unique genes passed the filtering were fed into multi-TGDR and Meta-TGDR algorithms.

*Statistical language and packages*

The statistical analysis was carried out in the R language version 2.15 (www.r-project.org), and packages were from the Bioconductor project (www.bioconductor.org). R code for multi-TGDR is available upon request.

**Results and Conclusions**

**Simulation studies**

In this section, we use two simulated examples to study the empirical performance of multi-TGDR.

**Example 1**

In the first simulation, 100×n iid standard normal (mean=0, variance=1) random variables (i.e., $X_1,…,X_{100}$, those are vectors of length n, n is the sample size), and n class membership outcome variables ($Y_1,…,Y_n$) taking the values of 1-3 were simulated. The logit function for class 2 and 3, having class 1 as reference, was calculated through the following relationship:

$$f_1 = 0.5 - 2X_1 + 1.2X_2 + 0.8X_3$$
$$f_2 = -1.5 + 1.7X_1 - 1.5X_2 - X_4$$

where the logit for class 2 depends only on features $X_1$ $X_2$ $X_3$ and class 3 logit depends on features 1,2 and 4. According to this model, 50 data sets were generated and analyzed by the proposed multi-TGDR framework. Results for this simulation, summarized in Table 1A shows that almost 100% of times the relevant features were selected by multi-TGDR framework. As criticized by Wang et al [21], lack of parsimony is an obvious disadvantage of TGDR algorithms, a shortcoming inherited by the multi-TGDR. However, the introduction of the Bagging procedure improves upon parsimony.

**Example 2**

To explore the effect of the correlations among features (i.e., independent variables) may have on multi-TGDR, we set the simulations as in the previous example but assumed the following correlations among features: cor($X_1$, $X_5$) =cor($X_3$, $X_7$)=0.8 and cor($X_2$, $X_6$) =cor($X_4$, $X_8$)=-0.8. Table 1B presents the results for this simulation. When compared with the uncorrelated scenario of example 1, the size of final selected feature is marginally larger while the predictive errors are almost the same. Nevertheless, multi-TGDR always selected the relevant features successfully and has good predictive performance. Bagging procedure improves upon both parsimony and predictive performance. Thus, it is highly recommended to combine bagging with any TGDR algorithm although bagging is very computing-time intensive.

**Applications on microarray expression data**

Using the real-world applications, the appropriateness and accuracy of the proposed multi-TGDR was evaluated. We also compared the performance of Meta-TGDR and TGDR under a priori batch-adjustment.

**Lung cancer**

About 80% of lung cancers (LC), the leading cause of cancer-related death throughout the world, are classified as non-small cell lung carcinoma (NSCLC) with Adenocarcinoma (AC) and squamous cell carcinoma (SCC) the two major subtypes of NSCLC. SCC is characterized as a poorly differentiated tumor subtype that develops in the proximal airways and is strongly associated with cigarette smoking. In contrast, AC usually arises in the peripheral airways and is more commonly observed in non-smokers and women. Mutations have been identified in AC and not in SCC, suggestion different mechanism of progression and treatment response.

For LC data, we randomly divided it into 4 subsets with roughly equal sizes and used 3 fold of them as the training set (n=109) and the remained 1-fold as the test set (n=36). Using multi-TGDR algorithm, 67 biomarkers were identified with 0% training error and a predictive error of 20.2% in a 5-fold cross-validation (CV). The comparison between pair-wise TGDRs and the multi-class TGDR is summarized in Table 2, and it shows that multi-TGDR over-perform the pair-wise strategy in all performance statistics.

After applying Bagging ($N_B$=100) to the LC data, we found that all genes in the 67-gene signature produced by multi-TGDR appear in the classifier with more than 5% frequency and 19 of them has bragging frequencies (BF) larger than 40% (Table 3). CYP24A1 is the gene most frequently selected (77%) followed by PNLDC1 (67%). By reducing the multi-TDGR signature to those genes with BF>40%, the performance showed slightly improvement by a predictive error reduction of 2.78% on the test set.

We concluded that bagging procedure discarded the random noises produced by a single run of TGDR. Furthermore, the calculation of membership probabilities in multi-class TGDR is more straightforward compared to the pair-wise coupling. Given it is intrinsically challenging to derive meaningful diagnostic signatures from high-throughput experiments in complicated problems like this, one major objective of presenting this data set is to use it as a benchmark for the development of more suitable classifiers on lung cancer subtypes and stages.

**Psoriasis**

Psoriasis vulgaris is a common chronic inflammatory skin disease of varying severity, characterized by red scaly plaques. Publicly data from 3 published studies [15-17] were used, including samples from Lesional (LS) and adjacent Non-Lesional (NL) skin of psoriasis patients and Normal skin from healthy patients.

Here, we used the psoriasis data to investigate the effect of the Batch/study adjustment in the performance of TGDR and Meta-TGDR, as well as to further evaluate multi-TGDR. Again, we randomly divided the whole data into 5 subsets with roughly equal sizes and used 4 fold of them as the training set (n=360) and the left one fold as one test set (n=89). By doing so, we can evaluate the validity of the proposed equation for the overall estimates since both the study-specific and overall estimates are available for the samples in this test set.

Here, we first present the performance of the binary classifiers followed by the multi-class problem. The binary classifiers (LS vs Normal, LS vs NL and NL vs Normal) will allow us 1) to assess the effect of batch adjustment on the performance of TGDR and Meta-TGDR and 2) to evaluate the validity of the method proposed here to allow prediction of independent data set in the Meta-TDGR framework.

*Binary Classification Problems in Psoriasis Data*

The results for all 3 comparisons were presented in Table 4A. The positive effect of batch adjustment on Meta-TDGR's performance is striking and consistent across all comparisons and datasets. When the data were adjusted for batch effect before classification, TGDR and Meta-TGDR had identical miss-classification rates on training and testing samples in all 3 comparisons and TGDR slightly outperformed in terms of GBS.

*LS versus normal:* Using TGDR on batch-adjusted expression from **LS and Normal** skin samples, we identified 30 biomarkers with a 0% and 0.86% training and CV-5 error, respectively. Bragging ($N_B$=100) frequencies were above 5% for all 30 genes. By considering a series of cutoff values for the frequency (5%-50%), a 30% for BF was chosen as it minimized the GBS and misclassification rates with the smallest number of non-zero genes leading to a final model with 18 genes (Table 4). Meta-TGDR (after batch adjustment) identified 22 biomarkers all with BF>5%. Cut-off for BF was set at 40% and among the 10 selected genes (see Table 5), 6 (p<0.0001) overlapped with the 18 genes in the TGDR bagging classifier.

*LS versus NL:* TGDR signature for **LS vs NL** classification included 35 biomarkers with a training error of 0% and 1.48% in 5-CV. Applying Bagging procedure ($N_B$=100) the final model (with BF>30%) included 22 genes (see Table 6). Meta-TGDR on the adjusted data identified a 25 genes signature, all with BF>5% and 16 of them above the selected cut-off of 40% for BF (See Table 6). There is still an impressive overlapping (n=11) between these 16 genes and the 22 genes chosen by TGDR bagging model (Fisher's test: p<0.0001).

Although Meta-TGDR is more parsimonious, k (the number of steps) is always dramatically bigger than in TGDR for the same value of the tuning parameter $\tau$ in both algorithms. One possible explanation is that by allowing different coefficients for a specific gene across studies, the direction of updating path in individual study may differ, probably leading to a cancel-out among one another. Therefore, the gradients might descend at higher speed in TGDR than meta-gradients in Meta-TGDR and the maximized likelihood value might be reached within fewer steps.

*Psoriasis 3-class classification*

Here we evaluate the performance of the multi-TGDR versus the classifier build using all pairwise binary TGDRs. The performance of the Meta-multi-TGDR for adjusted and unadjusted data is also presented.

Using multi-TGDR on the training set, we identified 60 genes with 0% training error 1.67% error in 5 fold CV. Interestingly, the number of selected genes by multi-TGDR is approximately the sum of all individualized binary TGDRs: with 39 genes (Table 6) having bagging frequency larger than the selected cut-off (40%). Again, disposal of the low-BF genes did not hurt the predictive performance. On the contrary, it improves the predictive performance in terms of GBS for both training and testing samples.

The classifier built by combining the 3 pairwise binary TGDR had the same in-training performance as multi-TGDR using 76 genes, with 44 of them being part of the multi-TGDR classifier (p<0.0001). The inconsistency between two algorithms is partially because local optimal points in individualized binary TGDRs cannot warrant the global optimality in the multi-TGDR.  With this data set, multi-TGDR and pairwise binary TGDR had similar performance while multi-TGDR was more parsimonious. Multi-TGDR with Bagging provided the best performance (see Table 7 and Figure 1). This shed some evidence on appropriation and accuracy of the multi-TGDR framework. Certainly, further evaluation using independent test sets is needed.

Surprisingly, the performance of Meta-multi-TGDR, where coefficients for both classes and studies are included is not impressive. This may partially due to the fact that Meta-multi-TGDR intends to find consistent-expressed genes across all classes and studies (one possible reason why the number of non-zero genes in multi-Meta-TGDR is the smallest).  Based on the analyses conducted here, we illustrated that TGDR on the adjusted data has a similar or better performance compared to Meta-TGDR, thus we think Meta-multi-TGDR, with its increase complexity and computing burden, is quite unnecessary. Nonetheless, the Meta-multi-TGDR greatly improved after batch adjustment reducing training error from 18.33 to 7.11 and predictive error (on test set 1) from 27% to 6.74, demonstrating that the adjustment of batch effect is imperative.

**Discussion**

When several microarray studies address the same or similar objectives, it is statistically more robust to carry out the analysis by pooling all studies together. To identify molecular signatures that discriminate among different disease status or stages on the pooled data, one can either apply TGDR to the batch-effect adjusted expression values for all samples, or use Meta-TGDR to select consistently informative genes and obtain the overall estimates using the procedure we proposed in the paper.

Using real-world applications, we showed that TGDR and Meta-TGDR have approximately equal predictive performance when the data has been adjusted for batch-effect. Compared to the latter method, TGDR on the adjusted data saves computing time, and do not require that all classes must be represented in each study. However, the stability of Meta-TGDR is usually better than TGDR as shown by the analyses of psoriasis data, and future work must be done to improve more on stability of TGDR. Nonetheless, applying Meta-TGDR on the unadjusted data had worse predictive performance compared to the analyses on the adjusted data. This verified our conjecture that Meta-TGDR aims mainly at selecting consistent genes across studies, with few to no capacity to adjust for a large batch-effect.

Additionally, we assembled our analyses with the Bagging procedure [12]. The benefits of Bagging including improved selection stability; more classification accuracy; and protection against over-fitting are clearly illustrated here.

In this paper, we did not compare the multi-TGDR with other classification methods. Since it is an extension to binary TGDR, whose performance had been proved to be equal or superior to many other classification methods in the original papers [10], we focus on important issues addressed here: comparing TGDR and Meta-TGDR performance after batch adjustment, making Meta-TGDR useful in practice by offering a solution to the prediction of independent datasets. Because the numbers of classes in our two applications are not big (4 in LC and 3 in psoriasis data, respectively), future work will include some applications of the multi-TGDR framework with a large number of classes, where the performance of pair-coupling has been reported to decrease dramatically, to see if a single likelihood-based classifier like multi-TGDR can be a rescue.

## Abbreviations

TGDR: threshold gradient descent regularization; Meta-TGDR: meta threshold gradient descent regularization; multi-TGDR: threshold gradient descent regularization for multiple classes; LC: lung cancer; NSCLC: non-small cell lung carcinoma; AC: adenocarcinoma; SCC: squamous cell carcinoma; DEGs: differentially expressed genes; LS: lesional; NL: non-lesional; CV-X: X fold cross validation; BF: bagging frequency; FDR: false discovery rate.

## Competing interests

None declared.

## Authors' Contributions

ST conceived the method, designed and implemented the Multi-TGDR algorithm, analyzed the data and wrote the manuscript. MSF analyzed the data, supervised the project, interpret the results and wrote of the manuscript. All authors reviewed and approved the final manuscript.

## Acknowledgements

This research was supported by a Clinical and Translational Science Award grant UL1RR024143; the National Science Foundation Grant No. PHYS-1066293 and the hospitality of the Aspen Center for Physics. MSF is also partially supported by the Milstein Program in Medical Research.

# Figures

**Figure 1. The estimated coefficients of the genes selected by multi-TGDR in the psoriasis data.**
Normal skin tissues from controls served as the reference. NL: Non-Lesional skin; LS: Lesional skin.

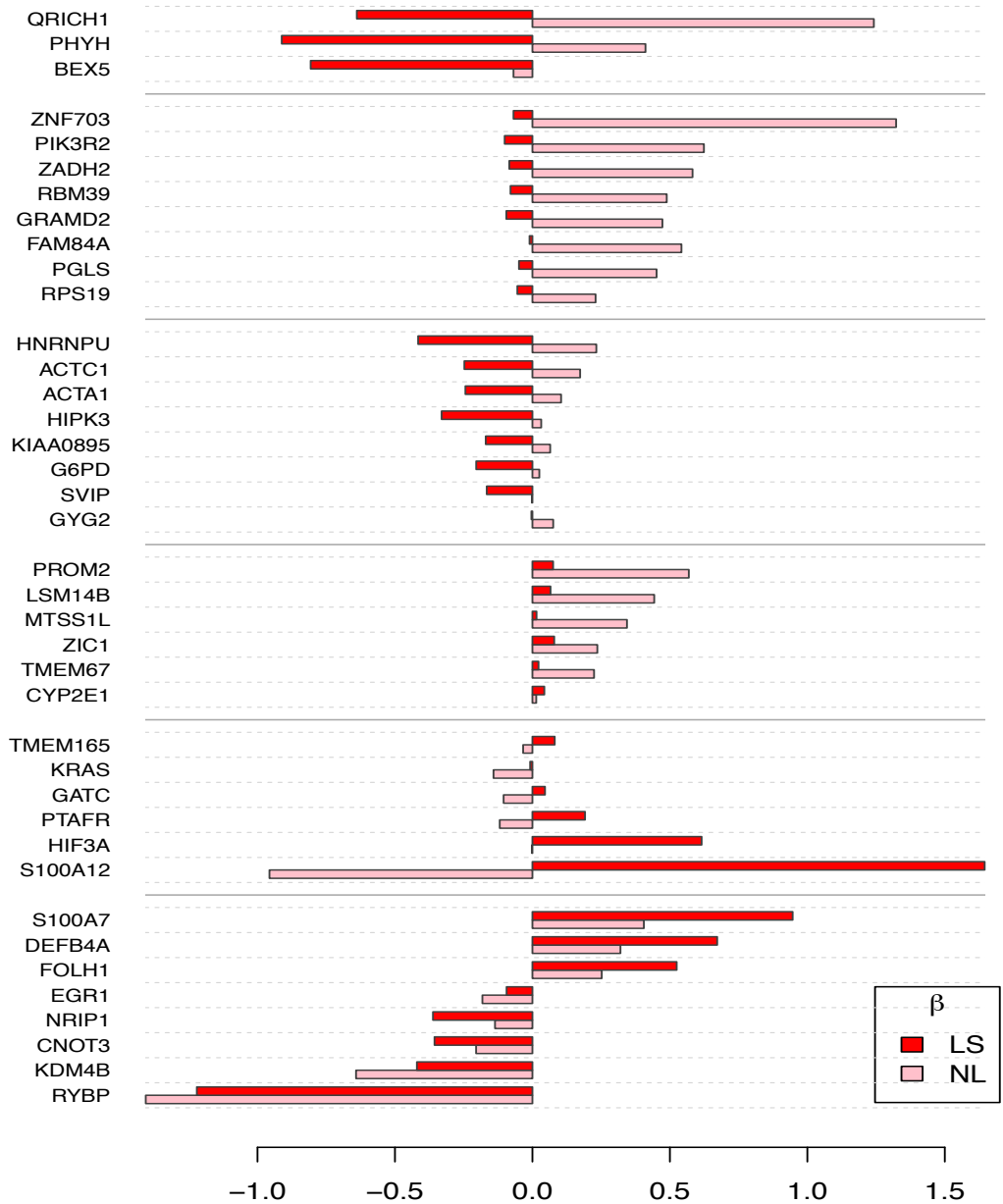

# Tables
## Table 1. The results for simulated data

| A. Simulation 1 | | | | | | |
|---|---|---|---|---|---|---|
| | % of $\beta_1\neq0$ Average BF (%) | % of $\beta_2\neq0$ BF(%) | % of $\beta_3\neq0$ BF(%) | % of $\beta_4\neq0$ BF(%) | Average # of selected genes | Average predictive error (%) |
| Multi-TGDR without bagging | 100 100 | 100 99.98 | 100 91.48 | 98 90.64 | 28.2 | 13.90 |
| Multi-TGDR BF>40% | --- | --- | --- | --- | 21 | 13.98 |
| BF>80% | --- | --- | --- | --- | 5.42 | 11.96 |
| B. Simulation 2 | | | | | | |
| Multi-TGDR without bagging | 100 100 | 100 100 | 96 90.88 | 100 95.48 | 30.20 | 13.26 |
| Multi-TGDR BF>40% | --- | --- | --- | --- | 22.62 | 13.20 |
| BF>80% | --- | --- | --- | --- | 5.54 | 11.08 |

## Table 2. Performance of classifiers for Lung Cancer data.

| | | Training (N=109) | | | Test set (n=36) | |
|---|---|---|---|---|---|---|
| | | # genes | Error (%) | CV (%) | Predictive Error (%) | GBS |
| Pair-wise | ACI vs ACII | 13 | 12.77 | 29.78 | | |
| | ACI vs SCC I | 28 | 0 | 11.63 | | |
| | ACI vs SCC II | 19 | 0 | 12.00 | | |
| | ACII vs SCCI | 15 | 0 | 3.39 | | |
| | ACII vs SCCII | 13 | 0 | 8.70 | | |
| | SCCI vs SCC II | 44 | 22.58 | 35.48 | | |
| | ***Overall*** | **107** | **18.34** | **51.38** | *50.00* | *0.302* |
| Multi-TDGR | *No Bagging* | **67** | ***0*** | **20.2%** | *47.22* | *0.292* |
| | with Bagging | 19 | 9.17 | --- | 44.44 | 0.303 |

## Table 3. Multi-TGDR genes for lung cancer data after Bagging.
Here, AC-I serves as the reference. Bagging frequency>40%.

| Probe | Symbol | Description | βAC-II | βSCC-I | βSCC-2 | Freq |
|---|---|---|---|---|---|---|
| 206504_at | CYP24A1 | cytochrome P450, family 24, subfamily A, polypeptide 1 | 0.2931 | -0.3501 | 0.0692 | 0.77 |
| 1564414_a_at | PNLDC1 | poly(A)-specific ribonuclease (PARN)-like domain containing 1 | -0.0499 | -0.0942 | 0.3569 | 0.67 |
| 211416_x_at | GGTLC1 | gamma-glutamyltransferase light chain 1 | -0.2747 | -0.1519 | -0.1726 | 0.65 |
| 206059_at | ZNF91 | zinc finger protein 91 | -0.0755 | 0.0557 | -0.2927 | 0.55 |
| 205348_s_at | DYNC1I1 | dynein, cytoplasmic 1, intermediate chain 1 | 0.1083 | 0.0092 | 0.388 | 0.54 |
| 231867_at | ODZ2 | odz, odd Oz/ten-m homolog 2 (Drosophila) | 0.0065 | 0.2616 | -0.0077 | 0.51 |
| 219926_at | POPDC3 | popeye domain containing 3 | 0.4104 | 0.1598 | 0.1239 | 0.51 |
| 219298_at | ECHDC3 | enoyl CoA hydratase domain containing 3 | -0.0718 | 0.0835 | -0.3062 | 0.5 |
| 203358_s_at | EZH2 | enhancer of zeste homolog 2 (Drosophila) | -0.5307 | -0.0307 | 0.4366 | 0.49 |
| 214464_at | CDC42BPA | CDC42 binding protein kinase alpha (DMPK-like) | -0.0406 | -0.3358 | -0.1537 | 0.48 |
| 210020_x_at | CALML3 | calmodulin-like 3 | -0.0172 | 0.2187 | 0.0875 | 0.46 |
| 201839_s_at | EPCAM | epithelial cell adhesion molecule | -0.0045 | -0.0408 | -0.0017 | 0.45 |
| 238983_at | NSUN7 | NOP2/Sun domain family, member 7 | 0.0135 | -0.2718 | -0.019 | 0.45 |
| 206677_at | KRT31 | keratin 31 | 0.0076 | -0.0138 | 0.1216 | 0.44 |
| 235706_at | CPM | carboxypeptidase M | 0.1549 | -0.0625 | 0.0164 | 0.43 |
| 226213_at | ERBB3 | v-erb-b2 erythroblastic leukemia viral oncogene homolog 3 (avian) | -0.0138 | -0.1827 | -0.0757 | 0.43 |
| 205713_s_at | COMP | cartilage oligomeric matrix protein | 0.0632 | -0.2135 | 0.048 | 0.41 |
| 228846_at | MXD1 | MAX dimerization protein 1 | 0.026 | 0.0184 | 0.0838 | 0.41 |
| 227492_at | OCLN | Occluding | 0.0015 | -0.4148 | -0.1417 | 0.41 |

**Table 4. Performance of Classifiers for Psoriasis data.** A. Comparison between TGDR and Meta-TGDR for binary classifiers. B. Comparisons between TGDR and Meta-TGDR for 3-class classifiers.

| A. | | Binary Classifiers | | | | | | |
|---|---|---|---|---|---|---|---|---|
| | | | | Training (n=360) | | | Test set (n=89) | |
| | | Method | # genes | Error (%) | 5-fold CV (%) | GBS | PE (%) | GBS |
| **LS vs Normal** | | TDGR (adjusted) | 30 | 0 | 0.86 | 0.0001 | 0 | 0.0006 |
| | | Meta-TGDR (unadjusted) | 18 | 0 | 0.86 | 0.0010 | 2.04 | 0.0084 |
| | | Meta-TGDR (adjusted) | 22 | 0 | 0.86 | 0.0006 | 0 | 0.0028 |
| Training: 233 | | TDGR w/Bagging (adjusted, BF>30%) | 18 | 0 | --- | 0.0011 | 0 | 0.0004 |
| Test: 49 | | Meta-TGDR w/Bagging (adjusted, BF>30%) | 10 | 0 | --- | 0.0012 | 0 | 0.0032 |
| **LS vs NL** | | TDGR (adjusted) | 35 | 0 | 1.48 | 0.0009 | 1.47 | 0.0136 |
| | | Meta-TGDR (unadjusted) | 26 | 1.11 | 1.85 | 0.0105 | 2.94 | 0.0294 |
| | | Meta-TGDR (adjusted) | 25 | 0 | 1.48 | 0.0036 | 1.47 | 0.0143 |
| Training: 271 | | TDGR w/Bagging (adjusted, BF>30%) | 22 | 0 | --- | 0.0021 | 1.47 | 0.0144 |
| Test: 68 | | Meta-TGDR w/Bagging (adjusted, BF>40%) | 16 | 1.48 | --- | 0.0041 | 1.47 | 0.0142 |
| **NL vs Normal** | | TDGR (adjusted) | 26 | 0 | 0 | $1.5 \times 10^{-5}$ | 0 | $7.3 \times 10^{-5}$ |
| | | Meta-TGDR (unadjusted) | 40 | 5.56 | 18.06 | 0.0570 | 8.20 | 0.0659 |
| | | Meta-TGDR (adjusted) | 22 | 0 | 1.85 | 0.0032 | 0 | 0.0054 |
| Training: 216 | | TDGR w/Bagging (adjusted, BF>30%) | 24 | 0 | --- | $2.4 \times 10^{-5}$ | 0 | $7.3 \times 10^{-5}$ |
| Test: 61 | | Meta-TGDR w/Bagging (adjusted, BF>40%) | 21 | 0 | --- | 0.0033 | 0 | 0.0054 |
| B. | | Multiclass Classifiers | | | | | | |
| Method | | | # genes | Error (%) | 5-fold CV (%) | GBS | PE (%) | GBS |
| Pairwise-coupling | | | 76 | 0 | 1.67 | 0.0067 | 0 | 0.0013 |
| Multi-TDGR | Adjusted | | 60 | 0 | 1.67 | 0.0144 | 0 | 0.0013 |
| | Bagging (BF>40%) | | 39 | 0 | --- | 0.0052 | 0 | 0.0006 |
| Multi-Meta-TGDR | Unadjusted | | 5 | 18.33 | 21.39 | 0.1734 | 26.97 | 0.1629 |
| | Adjusted | | 13 | 7.77 | 8.61 | 0.1296 | 6.74 | 0.0393 |

**Table 5 Psoriasis LS versus Normal genes by TGDR and Meta-TGDR after Bagging.** Here, Normal skin samples serve as the reference. Bagging frequency>30% for TGDR and >40% for Meta-TGDR.

| | | | TGDR | Meta-TGDR | | | |
|---|---|---|---|---|---|---|---|
| Probe | Symbol | Description | $\beta$ | $\beta^{Yao}$ | $\beta^{Gud}$ | $\beta^{SF+}$ | $\beta$ |
| 229963_at | BEX5 | brain expressed, X-linked 5 | -0.2958 | | | | |
| 207356_at | DEFB4A | defensin, beta 4A | 1.9258 | 1.3188 | 2.1196 | 1.6617 | 1.9405 |
| 224209_s_at | GDA | guanine deaminase | 0.8995 | 1.3512 | 1.4587 | 1.1601 | 1.5021 |
| 202411_at | IFI27 | interferon, alpha-inducible protein 27 | 0.69 | 0.0313 | 0.083 | 0.0556 | 0.2784 |
| 1555745_a_at | LYZ | lysozyme | 0.312 | 0.1929 | 0.0934 | 0.1837 | 0.0633 |
| 205916_at | S100A7 | S100 calcium binding protein A7 | 0.6612 | 0.2597 | 0.414 | 0.2862 | 0.4206 |
| 212492_s_at | KDM4B | lysine (K)-specific demethylase 4B | -0.1272 | | | | |
| 201846_s_at | RYBP | RING1 and YY1 binding protein | -1.4184 | -0.3202 | -0.1995 | -0.4103 | -0.2907 |
| 201416_at | SOX4 | SRY (sex determining region Y)-box 4 | -0.1703 | | | | |
| 215363_x_at | FOLH1 | folate hydrolase (prostate-specific membrane antigen) 1 | 0.3342 | | | | |
| 203335_at | PHYH | phytanoyl-CoA 2-hydroxylase | -0.3569 | | | | |
| 205758_at | CD8A | CD8a molecule | 0.1235 | | | | |
| 1556069_s_at | HIF3A | hypoxia inducible factor 3, alpha subunit | 0.2577 | | | | |
| 213424_at | KIAA0895 | KIAA0895 | -0.3158 | | | | |
| 205132_at | ACTC1 | actin, alpha, cardiac muscle 1 | -0.1815 | | | | |
| 1431_at | CYP2E1 | cytochrome P450, family 2, subfamily E, polypeptide 1 | 0.2671 | | | | |
| 230005_at | SVIP | small VCP/p97-interacting protein | -0.1723 | | | | |
| 202668_at | EFNB2 | ephrin-B2 | -0.1202 | | | | |

| Probe | Symbol | Description | | | | |
|---|---|---|---|---|---|---|
| 205471_s_at | DACH1 | dachshund homolog 1 (Drosophila) | -0.1171 | -0.0807 | -0.0722 | -0.0721 |
| 229625_at | GBP5 | guanylate binding protein 5 | 0.1256 | 0.0786 | 0.0659 | 0.1468 |
| 213293_s_at | TRIM22 | tripartite motif containing 22 | 0.1796 | 0.1428 | 0.1477 | 0.0042 |
| 202267_at | LAMC2 | laminin, gamma 2 | -0.0785 | -0.0844 | -0.0971 | -0.0942 |

**Table 6. Psoriasis LS versus NL genes by TGDR and Meta-TGDR after Bagging.** Non-lesional skin samples serve as the reference. Bagging frequency>30% for TGDR and >40% for Meta-TGDR.

| | | | TGDR | Meta-TGDR | | | |
|---|---|---|---|---|---|---|---|
| Probe | Symbol | Description | β | $β^{Yao}$ | $β^{Gud}$ | $β^{SF+}$ | β |
| 210002_at | GATA6 | GATA binding protein 6 | -0.1895 | | | | |
| 235603_at | HNRNPU | heterogeneous nuclear ribonucleoprotein U (scaffold attachment factor A) | -0.7306 | -0.4787 | -0.4382 | -0.5031 | -0.694 |
| 231875_at | KIF21A | kinesin family member 21A | -0.1396 | | | | |
| 233819_s_at | LTN1 | listerin E3 ubiquitin protein ligase 1 | -0.0771 | | | | |
| 203476_at | TPBG | trophoblast glycoprotein | 0.4798 | 0.1412 | 0.2202 | 0.2286 | 0.5236 |
| 234335_s_at | FAM84A | family with sequence similarity 84, member A | -0.1498 | | | | |
| 230828_at | GRAMD2 | GRAM domain containing 2 | -0.4782 | -0.0935 | -0.1705 | -0.306 | -0.1539 |
| 224171_at | LSM14B | LSM14B, SCD6 homolog B (S. cerevisiae) | -0.2053 | | | | |
| 230699_at | PGLS | 6-phosphogluconolactonase | -0.5561 | -0.0144 | -0.2076 | -0.328 | -0.1485 |
| 1552797_s_at | PROM2 | prominin 2 | -0.1398 | | | | |
| 226404_at | RBM39 | RNA binding motif protein 39 | -0.0434 | -0.1364 | -0.155 | -0.1385 | -0.2417 |
| 202648_at | RPS19 | ribosomal protein S19 | -0.4381 | -0.3294 | -0.6499 | -0.9706 | -0.7572 |
| 230586_s_at | ZNF703 | zinc finger protein 703 | -0.76 | -0.1284 | -0.3829 | -0.2787 | -0.9174 |
| 211661_x_at | PTAFR | platelet-activating factor receptor | 0.7216 | | | | |
| 203335_at | PHYH | phytanoyl-CoA 2-hydroxylase | -0.3971 | | | | |
| 213849_s_at | PPP2R2B | protein phosphatase 2, regulatory subunit B, beta | -0.1023 | | | | |
| 226367_at | KDM5A | lysine (K)-specific demethylase 5A | -0.3457 | -0.0369 | -0.0484 | -0.0447 | -0.0001 |
| 228132_at | ABLIM2 | actin binding LIM protein family, member 2 | -0.5047 | -0.5503 | -0.5072 | -0.5825 | -0.7550 |
| 202267_at | LAMC2 | laminin, gamma 2 | -0.1007 | -0.0953 | -0.0976 | -0.1271 | -0.0387 |
| 213424_at | KIAA0895 | KIAA0895 | -0.2031 | | | | |
| 205132_at | ACTC1 | actin, alpha, cardiac muscle 1 | -0.2045 | | | | |
| 203127_s_at | SPTLC2 | serine palmitoyltransferase, long chain base subunit 2 | 1.3511 | 0.4363 | 0.665 | 0.7786 | 0.9813 |
| 201487_at | CTSC | cathepsin C | | 0.0171 | 0.0155 | 0.0174 | 0.0000 |
| 217388_s_at | KYNU | Kynureninase | | 0.1287 | 0.1755 | 0.1958 | 0.0002 |
| 205863_at | S100A12 | S100 calcium binding protein A12 | | 0.3597 | 0.5641 | 0.5861 | 0.0486 |
| 243417_at | ZADH2 | zinc binding alcohol dehydrogenase domain containing 2 | | -0.1611 | -0.1703 | -0.1686 | -0.1063 |
| 211661_x_at | PTAFR | platelet-activating factor receptor | | 0.6241 | 0.6422 | 0.7737 | 0.9329 |

**Table 7. Psoriasis 3 classes genes selected by multi-TGDR after Bagging.** There are 39 genes in the final model. Normal tissue from healthy controls serves as the reference. Bagging frequency>40%.

| Probe | Symbol | Description | β_NL | β_LS | Freq |
|---|---|---|---|---|---|
| 203872_at | ACTA1 | actin, alpha 1, skeletal muscle | 0.1043 | -0.2438 | 0.51 |
| 229963_at | BEX5 | brain expressed, X-linked 5 | -0.0688 | -0.8066 | 0.75 |
| 207356_at | DEFB4A | defensin, beta 4A | 0.3202 | 0.672 | 0.46 |
| 235603_at | HNRNPU | heterogeneous nuclear ribonucleoprotein U (scaffold attachment factor A) | 0.2325 | -0.4157 | 0.68 |
| 205863_at | S100A12 | S100 calcium binding protein A12 | -0.9565 | 1.6449 | 0.42 |
| 205916_at | S100A7 | S100 calcium binding protein A7 | 0.4054 | 0.9465 | 0.81 |
| 226825_s_at | TMEM165 | transmembrane protein 165 | -0.0338 | 0.0811 | 0.58 |
| 206373_at | ZIC1 | Zic family member 1 | 0.2361 | 0.0792 | 0.59 |
| 203239_s_at | CNOT3 | CCR4-NOT transcription complex, subunit 3 | -0.2049 | -0.3561 | 0.43 |
| 201693_s_at | EGR1 | early growth response 1 | -0.182 | -0.094 | 0.47 |
| 234335_s_at | FAM84A | family with sequence similarity 84, member A | 0.542 | -0.0105 | 0.99 |
| 214711_at | GATC | glutamyl-tRNA(Gln) amidotransferase, subunit C homolog (bacterial) | -0.1048 | 0.0458 | 0.45 |
| 230828_at | GRAMD2 | GRAM domain containing 2 | 0.4728 | -0.0949 | 0.85 |
| 207764_s_at | HIPK3 | homeodomain interacting protein kinase 3 | 0.032 | -0.3303 | 0.54 |
| 212492_s_at | KDM4B | lysine (K)-specific demethylase 4B | -0.6416 | -0.4202 | 0.88 |
| 214352_s_at | KRAS | v-Ki-ras2 Kirsten rat sarcoma viral oncogene homolog | -0.1416 | -0.0083 | 0.55 |
| 224171_at | LSM14B | LSM14B, SCD6 homolog B (S. cerevisiae) | 0.443 | 0.0658 | 0.93 |
| 1556175_at | MTSS1L | metastasis suppressor 1-like | 0.3438 | 0.015 | 0.85 |
| 202600_s_at | NRIP1 | nuclear receptor interacting protein 1 | -0.1361 | -0.3617 | 0.61 |
| 230699_at | PGLS | 6-phosphogluconolactonase | 0.4517 | -0.049 | 0.79 |
| 229392_s_at | PIK3R2 | phosphoinositide-3-kinase, regulatory subunit 2 (beta) | 0.6238 | -0.101 | 0.78 |
| 1552797_s_at | PROM2 | prominin 2 | 0.5688 | 0.0752 | 0.84 |
| 229806_at | QRICH1 | glutamine-rich 1 | 1.2418 | -0.6387 | 0.86 |
| 226404_at | RBM39 | RNA binding motif protein 39 | 0.4885 | -0.0797 | 0.96 |
| 202648_at | RPS19 | ribosomal protein S19 | 0.2297 | -0.0548 | 0.9 |
| 201846_s_at | RYBP | RING1 and YY1 binding protein | -1.4064 | -1.2206 | 0.99 |
| 1563646_a_at | TMEM67 | transmembrane protein 67 | 0.2242 | 0.0222 | 0.48 |
| 243417_at | ZADH2 | zinc binding alcohol dehydrogenase domain containing 2 | 0.5824 | -0.0841 | 0.92 |
| 230586_s_at | ZNF703 | zinc finger protein 703 | 1.3228 | -0.0688 | 0.99 |
| 215363_x_at | FOLH1 | folate hydrolase (prostate-specific membrane antigen) 1 | 0.2521 | 0.5247 | 0.71 |
| 211661_x_at | PTAFR | platelet-activating factor receptor | -0.1189 | 0.1913 | 0.51 |
| 203335_at | PHYH | phytanoyl-CoA 2-hydroxylase | 0.4112 | -0.9118 | 0.61 |
| 202275_at | G6PD | glucose-6-phosphate dehydrogenase | 0.0251 | -0.2045 | 0.52 |
| 1556069_s_at | HIF3A | hypoxia inducible factor 3, alpha subunit | -0.0016 | 0.6154 | 0.62 |
| 213424_at | KIAA0895 | KIAA0895 | 0.0646 | -0.1698 | 0.48 |
| 215695_s_at | GYG2 | glycogenin 2 | 0.0756 | -0.0037 | 0.55 |
| 205132_at | ACTC1 | actin, alpha, cardiac muscle 1 | 0.1735 | -0.2476 | 0.6 |
| 1431_at | CYP2E1 | cytochrome P450, family 2, subfamily E, polypeptide 1 | 0.0139 | 0.0433 | 0.45 |
| 230005_at | SVIP | small VCP/p97-interacting protein | -0.0019 | -0.1661 | 0.43 |